\documentclass[10pt,oneside]{amsart}
\usepackage{adjustbox}

\usepackage{tikz}
\usetikzlibrary{automata, positioning, arrows}

      \usepackage{amssymb}
\usepackage{amsfonts,amstext}
\usepackage{graphicx}
\usepackage{url}
\usepackage{amsmath,amstext,amssymb,amscd}
\usepackage{verbatim} 
\usepackage{mathtools}
      \usepackage{tabularx,booktabs}
      \usepackage{multicol}

\usepackage{amssymb,latexsym}
\usepackage{amsmath}
\usepackage{amsthm}
\usepackage{amssymb}
\usepackage{makecell}
\newtheorem{thm}{Theorem}[section]

\theoremstyle{remark}

\theoremstyle{definition}

\newtheorem{example}[thm]{Example}
\newtheorem{remark}[thm]{Remark}

\usepackage{amsthm}
\allowdisplaybreaks

\usepackage{multicol}

      \makeatletter
      \def\@setcopyright{}
      \def\serieslogo@{}
     
      \makeatother

\begin{document}

\author{Mallory Dickerson}
\address{Mallory Dickerson, William Jewell College, 500 College Hill, Liberty, MO, 64068-1896}
\email{dickersonm.19@william.jewell.edu}

\author{Erin Martin}
\address{Erin Martin, Department of Mathematics and Data Science, William Jewell College, 500 College Hill, Liberty, MO, 64068-1896}
\email{martine@william.jewell.edu} 

\author{David McCune}
\address{David McCune, Department of Mathematics and Data Science, William Jewell College, 500 College Hill, Liberty, MO, 64068-1896}
\email{mccuned@william.jewell.edu}

\title[An Empirical Analysis of the Effect of Ballot Truncation]{An Empirical Analysis of the Effect of Ballot Truncation on Ranked-Choice Electoral Outcomes}

\begin{abstract}
In ranked-choice elections voters cast preference ballots which provide a voter's ranking of the candidates. The method of ranked-choice voting (RCV) chooses a winner by using voter preferences to simulate a series of runoff elections. Some jurisdictions which use RCV limit the number of candidates that voters can rank on the ballot, imposing what we term a \emph{truncation level}, which is the number of candidates that voters are allowed to rank. Given fixed voter preferences, the winner of the election can change if we impose different truncation levels.  We use a database of 1171 real-world ranked-choice elections to empirically analyze the potential effects of imposing different truncation levels in ranked-choice elections. Our general finding is that if the truncation level is at least three then restricting the number of candidates which can be ranked on the ballot rarely affects the election winner.
\end{abstract}



\maketitle

\section{Introduction}

The method of ranked-choice voting (also referred to as instant runoff voting, the Hare method, the plurality elimination rule, etc.) has been increasingly used in municipalities and states across the US since the turn of the century. The city of San Francisco began using the method for several municipal offices in 2004, followed by Minneapolis in 2009, Oakland in 2010, and several cities in Utah in 2021. New York City began using ranked-choice voting (RCV) for primary elections for city office in 2021; the Democratic primary election for mayor was the largest ranked-choice election ever held in the US, with almost one million votes cast. RCV has also been used for elections for federal office in Maine (since 2018) and Alaska (since 2022). In ranked-choice elections, voters cast a preference ballot which provides a preference ranking of the candidates. Often, voters do not provide a full ranking of the candidates, instead casting a \emph{partial ballot} in which only a partial ranking of the candidates is provided.  Partial ballots arise for two reasons. First, voters may choose not to provide a complete ranking, deciding to leave some candidates off their ballots. Second, some jurisdictions limit the number of candidates that voters can rank on their ballots. For example, Minneapolis (respectively New York City) allows a voter to rank only three (respectively five) candidates on their ballots, regardless of how many candidates are in the race. We say that an election's \emph{truncation level} is the number of candidates which can be ranked on a ballot, so that the truncation level of Minneapolis municipal elections is three, for example. The purpose of this article is to investigate empirically how the choice of truncation level can affect the outcome of a ranked-choice election. 

There are two main ways in which an electoral outcome can be affected when a jurisdiction imposes a truncation level:

\begin{enumerate}
\item In response to the truncation level, voters may choose to provide a different ranking of their top candidates than they otherwise would if the limit did not exist. For example, in Minneapolis a voter might not rank their three favorite candidates on their ballot because the voter might be worried that none of these candidates have a legitimate chance of winning, and thus the voter might choose to rank three ``stronger'' but less preferred candidates instead.
\item The RCV winner of an election without an imposed truncation level can be different from the winner when voters can provide a complete ranking, even if voters do not change how they rank the top candidates on their ballot in response to the limit.
\end{enumerate}

It is very difficult to investigate the effect of (1) on electoral outcomes because we cannot know how voters would have voted if the limit were removed. However, (2) can be investigated in a variety of ways. The most thorough examination can be found in \cite{KGF}, which primarily used Monte Carlo simulation to investigate how the RCV winner can change depending on the choice of truncation level, assuming voters do not change their preferences as we vary that limit. \cite{KGF} also used real-world data in their analysis, but their empirical work included only 18 elections, and only 6 of those were from real-world elections for political office (which is our primary focus). We build on their work by using 1171 ranked-choice elections, 1148 of which are political elections from the US and Scotland. Thus, this article can be read as an empirical companion piece to \cite{KGF}, and our work is by far the largest empirical study to date of the potential effects of ballot truncation on electoral outcomes under RCV.

The issue of partial ballots and the resulting potential effects on ranked-choice elections has been studied from many angles. \cite{BK} analyzes four ranked-choice elections in which the RCV winner did not obtain a majority of the vote because a large proportion of the ballots were partial. Several studies (\cite{BFLR},  \cite{CLMM}, \cite{KL}) study the problem of which candidates could become the RCV winner if partial ballots were filled in to create full rankings. Other studies (\cite{BFLR}, \cite{FB}, \cite{K}) consider how partial ballots can be used for strategic voting or strategic campaigning, focusing on issues such as truncation or no-show paradoxes. Our work is most similar to \cite{TUK}, which contains an empirical component addressing the issue of how many different candidates could be the RCV winner as the truncation level varies. What sets our empirical work apart from theirs is that we use a much larger database of elections and we include an analysis of the effect of the truncation level on the likelihood of electing the Condorcet winner.

\section{Preliminaries}\label{prelims}

Let $n$ be the number of candidates in an election and let $TL$ denote the election's \emph{truncation level}, the number of candidates that voters are allowed to rank on their ballots. In a ranked-choice election, voters submit preference ballots, providing a ranking of up to $TL$ candidates on their ballots. The ballots are then aggregated into a \emph{preference profile}, which shows how many ballots of each type were cast. Table \ref{preference_profile} provides an example of a preference profile for an election in which $n=TL=4$, and the candidates are labeled $A, B, C,$ and $D$. The number 57 denotes that 57 voters cast a ballot ranking $A$ first, $B$ second, and no candidate ranked third or fourth; the other numbers across the top row convey similar information about the number of ballots cast. For all of the elections in our database voters are not required to provide a complete ranking, and the preference profile in Table \ref{preference_profile} has many voters who choose not to rank all four candidates.

To choose a winner of an election given a preference profile, RCV proceeds in a series of rounds. In each round, a candidate's first-place votes are counted; a candidate with a majority of the votes is declared the winner. If no candidate achieves  a majority then the candidate with the fewest first-place votes is eliminated and their votes are transferred to the next candidate on their ballots who has not previously been eliminated. The method continues in this fashion until a candidate achieves a majority of the remaining votes. We note this method can be extended in a variety of ways to elect multiple candidates given a preference profile; in this article we focus only on the single-winner case.

\begin{table}[]
  \centering

\begin{tabular}{l|c|c|c|c|c|c|c|c|c|c|c}
Num. Voters&57&26&51&137&2&38&16&53&72&15&33\\
\hline
1st choice&$A$&$A$&$A$&$B$&$B$&$C$&$C$&$C$&$D$&$D$&$D$\\
2nd choice&$B$&$D$&$D$&$C$&$C$&$A$&$B$&$D$&$A$&$A$&$B$\\
3rd choice& && $C$     &$A$&$D$&$D$&$D$&$A$&$B$&$C$&$A$ \\
4th choice&       &&     $B$  &    &      &       &    &  $B$    &  &$B$&$C$    \\

\end{tabular}

\caption{A preference profile with four candidates.}
  \label{preference_profile}
\end{table}

We illustrate the RCV algorithm using the preference profile from Table \ref{preference_profile}, where we assume $TL=4$ so that none of the voters' preferences are truncated.

\begin{example}\label{first_ex}
Initially, the number of first-place votes for $A$, $B$, $C$, and $D$ are 134, 139, 107, and 120, respectively. No candidate achieves a majority of first-place votes, and thus $C$ is eliminated. Consequently 38 votes are transferred to $A$, 16 are transferred to $B$, and 53 are transferred to $C$, resulting in adjusted vote totals of 172, 155, and 173 for $A$, $B$, and $D$, respectively. There is still no candidate with a majority and thus $B$ is eliminated. Since $C$ has already been eliminated, 137 votes are transferred to $A$ and 2 are transferred to $D$, and $A$ wins the election with 309 votes to $D$'s 191.
\end{example}

We now illustrate how the use of a truncation level can affect the winner of a ranked-choice election.

\begin{table}[]
  \centering

\begin{tabular}{l|c|c|c|c|c|c|c|c|c|c|c}
\multicolumn{12}{c}{$TL=1$, RCV winner $= B$}\\
\hline
\hline
Num. Voters&57&26&51&137&2&38&16&53&72&15&33\\
\hline
1st choice&$A$&$A$&$A$&$B$&$B$&$C$&$C$&$C$&$D$&$D$&$D$\\

\end{tabular}
\vspace{.1in }

\begin{tabular}{l|c|c|c|c|c|c|c|c|c|c|c}
\multicolumn{12}{c}{$TL=2$, RCV winner $= D$}\\
\hline
\hline
Num. Voters&57&26&51&137&2&38&16&53&72&15&33\\
\hline
1st choice&$A$&$A$&$A$&$B$&$B$&$C$&$C$&$C$&$D$&$D$&$D$\\
2nd choice&$B$&$D$&$D$&$C$&$C$&$A$&$B$&$D$&$A$&$A$&$B$\\
\end{tabular}
\vspace{.1 in}

\begin{tabular}{l|c|c|c|c|c|c|c|c|c|c|c}
\multicolumn{12}{c}{$TL=3$, RCV winner $=A$}\\
\hline
\hline
Num. Voters&57&26&51&137&2&38&16&53&72&15&33\\
\hline
1st choice&$A$&$A$&$A$&$B$&$B$&$C$&$C$&$C$&$D$&$D$&$D$\\
2nd choice&$B$&$D$&$D$&$C$&$C$&$A$&$B$&$D$&$A$&$A$&$B$\\
3rd choice& && $C$     &$A$&$D$&$D$&$D$&$A$&$B$&$C$&$A$ \\

\end{tabular}

\caption{The preference profile from Table \ref{preference_profile} with truncation levels 1, 2, and 3.}
  \label{preference_profiles_trunc}
\end{table}

\begin{example}\label{second_ex}
Table \ref{preference_profiles_trunc} shows preference profiles we obtain from the profile in Table \ref{preference_profile} for truncation levels of 1, 2, and 3. When $TL=1$, the RCV algorithm merely selects the candidate with the most first-place votes (referred to as the election's \emph{plurality winner}), which is $B$ with 139 first-place votes. 

When $TL=2$, $C$ is eliminated first, resulting in vote totals of 172, 155, and 173 for $A$, $B$, and $D$, respectively, just as in Example \ref{first_ex}. However, in this case when $B$ is eliminated $D$ defeats $A$ with 173 votes to $A$'s 172. In the original example with $TL=4$, $A$ was able to defeat $D$ because $A$ received 137 votes from the elimination of $B$; this is no longer the case as $A$ was these 137 voters' third choice, and when $TL=2$ we do not register voters' third choices.

When $TL=3$, the RCV algorithm proceeds exactly as the original example where $TL=4$, and $A$ wins the election.
\end{example}

The previous example leads to two remarks, which have been pointed out by many others (see \cite{KGF} and \cite{TUK}, for example). 

\begin{remark}\label{obs_1}
In an $n$-candidate election, the RCV winner when $TL=n$ is the RCV winner when $TL=n-1$.
\end{remark}

\begin{remark}\label{obs_2}
In an $n$-candidate election, as we vary the truncation level from 1 to $n$ the number of different RCV winners is at most $n-1$. 
\end{remark}

Example \ref{second_ex} demonstrates Remark \ref{obs_2}: depending on the truncation level, the RCV winner can be $A$, $B$, or $D$. By Remark \ref{obs_1}, it is not possible for the fourth candidate $C$ to be an RCV winner as well under some truncation level.  \cite{TUK} show that the bound in Remark \ref{obs_2} is sharp, in that for any $n$ there exists an $n$-candidate election with $n-1$ different RCV winners as $TL$ is varied from 1 to $n-1$.


In addition to analyzing the effect of ballot truncation on the RCV winner, \cite{KGF} investigates how the truncation level affects the likelihood that RCV elects the \emph{Condorcet winner} of the election. The concept of a Condorcet winner dates back to the social choice debates between Jean-Charles de Borda and the Marquis de Condorcet (see \cite{B} for an interesting exposition of this history).  A candidate is a Condorcet winner if this candidate beats all other candidates in head-to-head matchups;  such candidates receive much attention in the social choice literature because they are seen as ``strong'' or ``deserving'' candidates. To provide an example, note that for the election in Table \ref{preference_profile} candidate $A$ defeats $B$ in a head-to-head matchup because $57+26+51+38+53+72+15=312$ voters prefer $A$ to $B$, while only 188 voters prefer $B$ to $A$. Similarly, 254 voters prefer $A$ to $C$ while 246 prefer $C$ to $A$, and 309 voters prefer $A$ to $D$ while 191 prefer $D$ to $A$. Thus, $A$ is the Condorcet winner of this election. Example \ref{second_ex} shows that the choice of truncation level can effect whether the Condorcet winner wins the election under RCV.

In our description of the Condorcet winner of the election in Table \ref{preference_profile} we assumed that the candidates left off a voter's ballot are all tied for last place in that voter's preferences; that is, we choose to process partial rankings using the \emph{weak order model} \cite{PPR}. There are other ways to process partial ballots which would potentially change our analysis, but we prefer the weak order model because it deals with exactly the information provided by the voters. Furthermore, election offices which run RCV elections process ballots under this model, although they do not use that language.

To conclude this section, we list the questions in which we are interested, which we use real-world ranked-choice data to investigate.

\begin{itemize}
\item \textbf{Question 1}: For a fixed truncation level $TL$, what percentage of elections satisfying $TL<n-1$ have the property that the RCV winner when using $TL$ is different from the RCV winner when preferences are not truncated?
\item \textbf{Question 2}: For $1\le k\le n-1$, what percentage of elections have $k$ different winners as we increase $TL$ from 1 to $n-1$?
\item \textbf{Question 3}: As we increase the truncation level, does the likelihood of electing the Condorcet winner increase? 
\end{itemize}

Before providing our results, we describe our data.

\section{Data Sources}

Our database of 1171 elections comes from three sources. We briefly describe each source and the number of elections we collected from it. Note that if an election contains a candidate who achieves a majority in the first round then it is not interesting to investigate the effects of truncation; furthermore, to see the full effects of truncation, we are interested only in elections in which voters can express a complete ranking of the candidates if they so choose. Thus, our database consists of elections in which the RCV algorithm goes to at least a second round and the truncation level used in the actual election satisfies $TL\ge n-1$.

\textbf{American Psychological Association}: The American Psychological Association (APA) uses RCV to elect its President and positions on its Board of Directors. Our database includes APA presidential elections from the years 1998-2009 and 2017-2021, as well as four elections for Board of Director positions from the years 2017-2021. These elections were collected by the third author for \cite{MMa}. The elections from 1998-2009 are available at \cite{MW}, and the 2017-2021 elections were provided directly by the APA. As far as we are aware, data from 2010-2016 is not available. Our database includes 23 elections from the APA, all of which are all single-winner.

\textbf{American Ranked-Choice Political Elections}: As mentioned in the introduction, many American jurisdictions have started using RCV. Hundreds of RCV elections have occurred across the US since 2004, mostly for municipal office such as mayor or city council positions. Many of these elections are not useable for our purposes because they contain a majority candidate or they satisfy $TL<n-1$. Our database includes 82 American political elections, 80 of which are single-winner. These elections were collected by the third author for \cite{MMa}, and most of them are now publicly available at the FairVote data repository \cite{O2}.

\textbf{Scottish Local Government Elections}: For the purposes of local government, Scotland is divided into 32 council areas, each of which is overseen by a council. These council areas are roughly analogous to counties in the US, and the councils provide a range of services that Americans associate with services provided by county or municipal governments. Each council area is divided into wards, and every five years councilors from each ward are elected to represent the ward on the council using a multiwinner version of RCV called \emph{single-transferable vote}. Scottish wards have used this version of RCV since 2007. Data from that year is essentially unavailable; most of the vote data for the 2012 and 2017 election cycles is available at \cite{T}. The third author collected data for the 2022 elections from various council election offices for \cite{MGS}. Sometimes by-elections are held in off-cycle years to fill a seat left vacant by a death or resignation; the third author collected the ballot data for as many of these by-elections as possible directly from council election offices (often, this data was not publicly posted).

Of the Scottish elections 29 are single-winner and 1037 are multiwinner. We acknowledge it is not optimal that so many of our elections are multiwinner, since we study the effects of truncation only for RCV in the single-winner case. However, these multiwinner Scottish elections provide preference data for candidates in political elections, and thus are still of use for our purposes. In particular, these multiwinner elections tend to produce very close results from a single-winner perspective, and analyzing them could help provide an upper bound for the frequency with which truncation affects electoral outcomes (we expect truncation to matter infrequently if elections are not ``close'' in some sense; the most extreme example of an election which is not close is an election with a majority candidate, where the truncation level is irrelevant).

In all, 132 of our elections are single-winner and 1039 are multiwinner. Table \ref{num_cands} summarizes the database by the number of candidates in each election, where the number of candidates does not include write-ins. Given a fixed truncation level, the question of whether the RCV winner would differ from the RCV winner in the truncated election is interesting only if $TL< n-1$, and thus the table also shows how many elections are of interest for a fixed truncation level (we refer to such elections as \emph{useable}). For example, if we use $TL=3$ as Minneapolis does, then our database contains 1073 elections (corresponding to $n\ge 5$) in which the RCV winner with truncation might differ from the RCV winner without.

\begin{table}[]
  \centering

\begin{tabular}{l|cccccccccccc|c}
$n$    & 3&4&5&6&7&8&9&10&11&12&13&14&Total\\
\hline
APA &6 & & 17 &&&&&&&&&&23\\ 
Amer.&24& 31& 13 &8&4&&1&1&&&&&82\\ 
 Scot. & 2&35&117&210&286&205&112&63&22&8&5&1&1066\\
\hline
Total & 32 & 66 & 147 & 218 & 290 & 205 & 113 & 64 & 22 & 8 & 5 & 1&1171\\
\end{tabular}
\caption{The number of elections with a given number of candidates from each data source.}
  \label{num_cands}
  \end{table}
  
  As mentioned in the introduction, the size of our database is partly what sets our work apart from previous studies. The largest prior empirical study of the effect of ballot truncation on the RCV winner is \cite{TUK}, which analyzes 168 ranked-choice elections (the ballot data for these elections is available from \cite{MW}). Our database includes some of these elections but, in our view, many of the elections analyzed in \cite{TUK} are not ideal for this analysis. Some of their elections contain majority candidates and in some voters were not allowed to provide a complete ranking. In addition to the size of our database, we also include a detailed analysis of the effect of truncation on the Condorcet winner, which has only been done previously by \cite{KGF} in 18 elections.

 We conclude this section by noting that real-world data is often messy and presents challenges beyond the choice of how to process partial ballots. For example, voters sometimes give two candidates the same ranking, resulting in a ballot error known as an \emph{overvote}. When this occurs, most election offices in the US choose to cut off the voter's ballot at the ranking where the overvote occurs, ignoring the candidate choices at that ranking and any subsequent rankings. Also, sometimes voters skip multiple rankings on their ballots, perhaps casting a ballot in which $A$ is ranked first and $D$ is ranked fourth, but no candidates are ranked second or third. Most election offices in the US interpret this ballot as saying that $A$ is the voter's first choice and $D$ is the second choice, just ignoring the skipped rankings, but a few election offices disregard any rankings on a ballot which occur after two skipped rankings. When we process ballots we choose to process them in line with how we think the given jurisdiction processes them (although sometimes there is ambiguity). Fortunately, most of our data avoids these issues, as the APA and Scottish elections are provided to us in a clean fashion; the APA and the Scottish election offices deal with the aforementioned issues before sharing the data, essentially providing their data in the form of Table \ref{preference_profile}. Thus, we have to deal with issues of voter error only in the 82 American political elections.

\section{Results}
We now present our results, using our elections database to answer Questions 1-3 from Section \ref{prelims}. As mentioned in the introduction, our results are meant to be read as an empirical complement to the results of \cite{KGF}, who mainly use Monte Carlo simulations based on a spatial and random model of voter preferences to provide theoretical results. Thus, we contrast our findings with theirs as part of our discussion, although \cite{KGF} study only elections with $n\in \{4,5,6\}$.

\textbf{Question 1}: For a fixed truncation level $TL$, what percentage of elections satisfying $TL<n-1$ have the property that the RCV winner when using $TL$ is different from the RCV winner when preferences are not truncated?

Table \ref{Q1results} shows, for a fixed $TL$, the percentage of elections in which the RCV winner without truncation is not the RCV winner with the truncation level imposed. For $TL=1$, we see that the percentage of elections in which the candidate who receives the most first-place votes (the election's plurality winner) is also the RCV winner without truncation in 76.1\% of the elections in the database. Thus, for more than three-quarters of our elections the RCV algorithm is in some sense unnecessary, as the algorithm simply chooses the plurality winner and any preference information past the first ranking does not make a  difference.

Note that when $TL=3$, which is the case for elections in municipalities such as Minneapolis, the two winners with and without truncation agree in 96.8\% of the elections, suggesting that truncating at depth 3 almost never makes a difference as long as voters do not adjust their stated preferences in response to the imposed truncation level. Furthermore, we did not include all possible truncation levels in order to keep our table a manageable size, but we note that there is one interesting Scottish election (which in reality was multiwinner) with 11 candidates in which the winner at $TL=9$ is different from the winner with no truncation. In the 2012 Ward 3 council election of the North Ayrshire council area, candidate John Ferguson is the RCV winner for all truncation levels $1\le TL \le 10$ except $TL=9$, where Joe Cullinane is the winner. This is the only election in which we observe a difference in winners for $TL>5$.

As often occurs when comparing empirical to theoretical findings, our percentages are much lower than what is reported by \cite{KGF} under either of their models. For example, under their random model the probabilities that the winner with no truncation is different from the winner with $TL=2$ in 4, 5, and 6 candidate elections are approximately 20\%, 30\%, and 40\%, respectively. The probabilities for $TL=2$ are similar under their spatial model. Theoretical models are often used to establish upper bounds for the probabilities of various voting phenomena, and empirical work investigates how far the real-world data is from these theoretical upper bounds. In this case, what we see in practice is significantly different than what was produced by the theoretical models.

\begin{table}[]
  \centering

\begin{tabular}{l|c|c|c|c|c|c|c}
$TL$&1&2&3&4&5&6&7\\
\hline
Useable elections &1171 & 1139 & 1073&926&708&418&213\\
\% Winners Agree&76.1\% & 89.7\% & 96.8\% & 98.9\% & 99.7\% & 100\% & 100\%\\
\end{tabular}
\caption{For a given $TL$, the percentage of elections in which the RCV winner without truncation is the winner with the imposed truncation level.}
  \label{Q1results}
  \end{table}

\textbf{Question 2}: For $1\le k\le n-1$, what percentage of elections have $k$ different winners as we increase $TL$ from 1 to $n-1$?

We do not find any elections which demonstrate more than three different winners as we vary the $TL$ from 1 to $n-1$. The database contains 14 elections which demonstrate three different winners; all of these elections are from the set of multiwinner Scottish elections. These elections account for only 1.5\% of the elections satisfying $n\ge 4$. The database contains 293 elections which demonstrate two different winners, accounting for 25.0\% of the database.  73.8\% of the elections produce the same RCV winner regardless of the imposed truncation level.

\begin{figure}[!htb] 
\begin{center}
\begin{tabular}{c}
\includegraphics[scale=0.45]{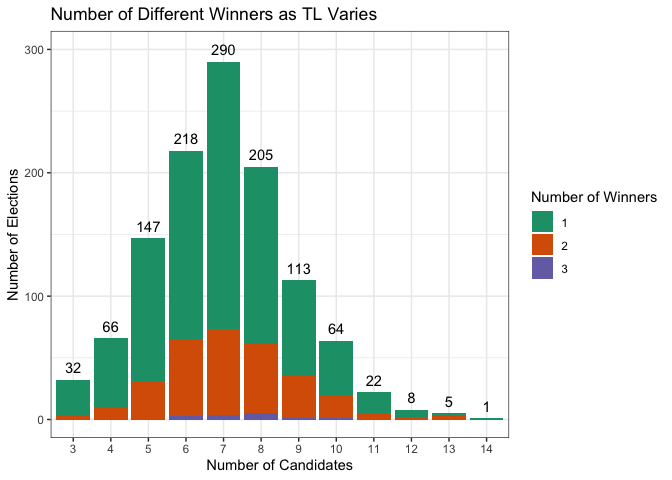}\\
\\
 \includegraphics[scale=0.45]{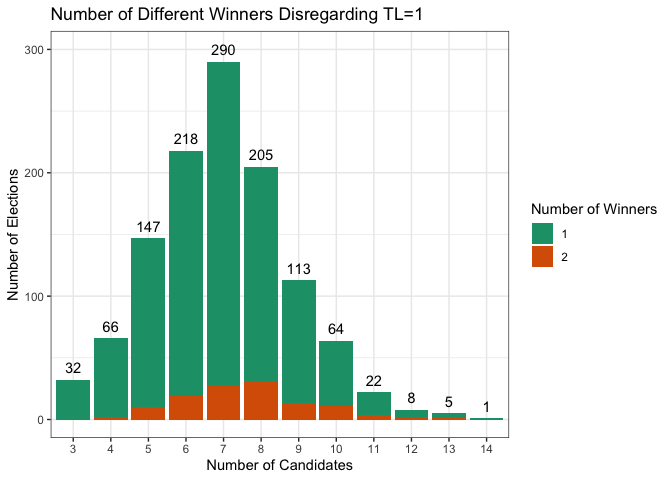} \\

\end{tabular}

\end{center}
\caption{(Top) The number of different winners in each election as $TL$ varies from 1 to $n-1$, separated by the number of candidates. (Bottom) The number of different winners in each election as $TL$ varies from 2 to $n-1$.}
\label{first_figure}
\end{figure}

The top image of Figure \ref{first_figure} shows these results broken down by the number of candidates. For example, the figure shows that of the 218 6-candidate elections in the database, three elections produce three different winners as $TL$ varies from 1 to 5, 62 produce two different winners, and 153 produce only one winner. Complete details about these results are available at \cite{G}. Recall from Remark \ref{obs_2} that the maximum number of different winners is $n-1$; we do not observe any such extreme outcomes in our data except for the uninteresting case when $n=3$. The elections with 3 different winners come from elections satisfying $n\ge 6$. Furthermore, when an election produces more than one winner, often this is because the plurality winner is different from the RCV winner at all other truncation levels. The bottom image of Figure \ref{first_figure} shows the number of different winners broken down by number of candidates, where we use only truncation levels satisfying $2\le TL \le n-1$. We expect the plurality winner to differ from the RCV winner in a significant number of elections because the calculation of a plurality winner does not require preference data, and thus it is not particularly interesting if the plurality winner causes an increase in the total number of different winners. Figure \ref{first_figure} shows that the handful of elections which produced three different winners all achieved this outcome by having the plurality winner be different from the winners at all other truncation levels. If we use any preference data from the ballots past the first ranking, the maximum number of different winners we see is only two.

\begin{figure}[tbh] 
\begin{center}
\includegraphics[scale=0.45]{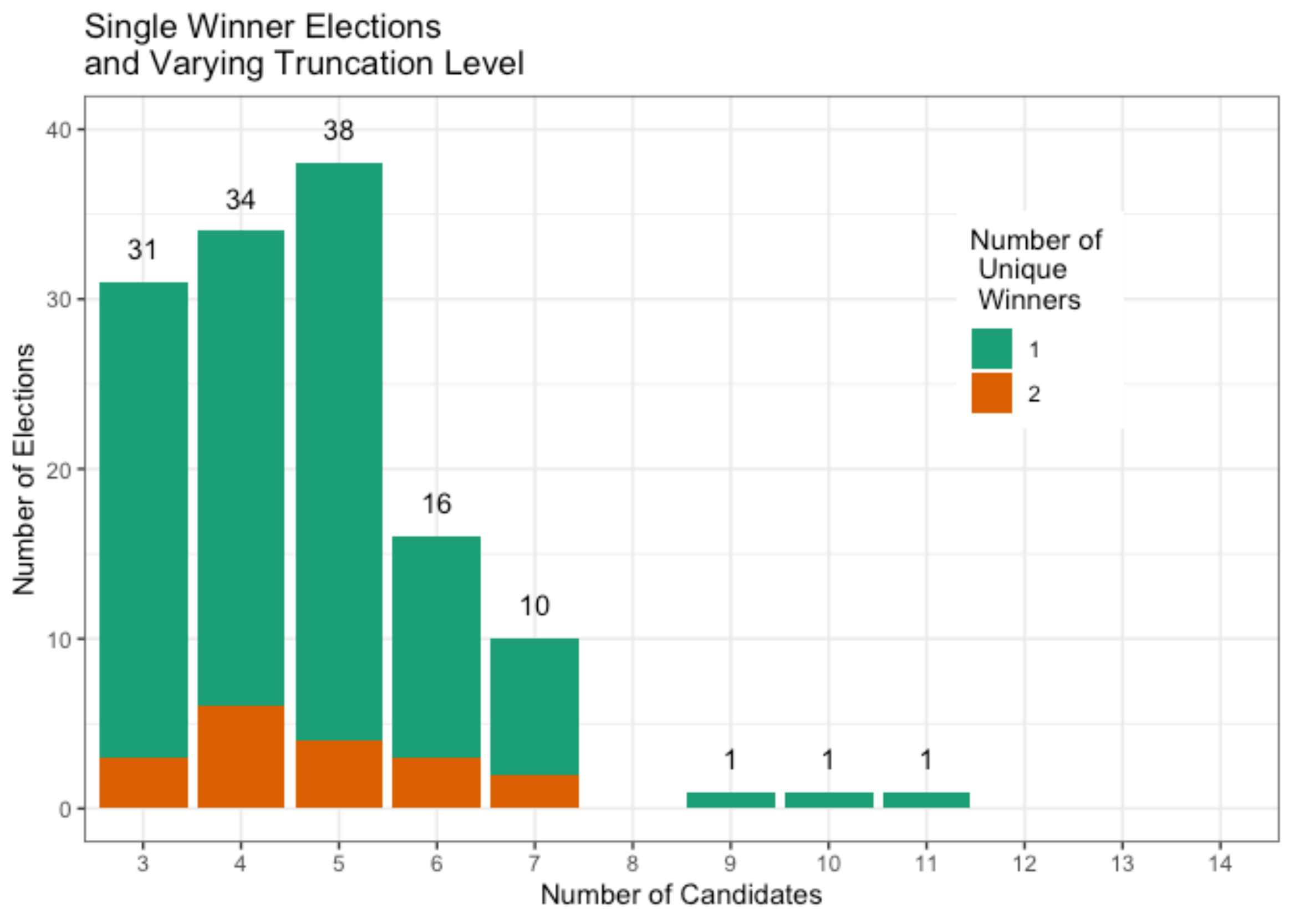}\\

\end{center}
\caption{The number of different winners in each of the 132 single-winner elections as $TL$ varies from 1 to $n-1$.}
\label{singlewinner_figure}
\end{figure}

Figure \ref{singlewinner_figure} is the version of the top image of Figure \ref{first_figure} where we use only the 132 single-winner elections in the database. For these elections we see at most two different winners. Furthermore, most of the elections which produce two winners are ``uninteresting'' in the sense that we get two winners only because the plurality winner is different from the winner at all other truncation levels. There are only three of these elections in which the RCV winner at a truncation level greater than 1 is different from the actual RCV winner. We list these elections below.

\begin{itemize}
\item The June 2021 Republican primary election for the District 50 city council seat in New York City. Mario Kepi wins for $TL \in \{1, 2\}$ and David Carr wins for $TL \in \{3,4\}$.

\item The 2021 Scottish by-election in the Isle of Bute ward of the Argyll and Bute council area. Kim Findlay wins for $TL \in \{1,2,3\}$ and Liz McCabe wins for $TL=4$.

\item The 2021 Scottish by-election 2021 in the Patrick East/Kelvindale ward of the Glasgow City council area. Abdul Bostani wins for $TL \in \{1,2\}$ and Jill Brown wins for $TL \in \{3,4,5\}$.

\end{itemize}

As was the case with Question 1, for this question we find much lower levels of winner disagreement than do \cite{KGF}. For example, under their spatial model, in 6-candidate elections their simulations return estimated approximate probabilities of 45\%, 55\%, and 5\% that an election produces 1, 2, or 3 different winners, respectively. Their simulations also returned a handful of 6-candidate elections with 4 different winners. From Figure \ref{first_figure} we see that these relatively large disagreement probabilities are not found in our data for any number of candidates.

\textbf{Question 3}: As we increase the truncation level, does the likelihood of electing the Condorcet winner increase? 

 To investigate the effects of truncation on the likelihood of electing the Condorcet winner, only some of the elections in the database are relevant for our analysis. We say that an election is \emph{Condorcet-useable} if the election contains a Condorcet winner and the Condorcet winner is the RCV winner for some truncation level. If an election is not Condorcet-useable then it is not useful for investigating the effect of truncation on whether the Condorcet winner is the RCV winner. Of the 1171 elections, 13 do not contain a Condorcet winner and 52 have a Condorcet winner but this candidate is not the RCV winner for any truncation level, resulting in 1106 Condorcet-useable elections.
 
 The Condorcet results are given in Table \ref{Cond_results}. The table  shows that, in general, a higher truncation level leads to an increased likelihood that the Condorcet winner will win the election. Furthermore, it seems that the increased likelihood tends to level off at approximately 98.5\%, and this convergence occurs by $TL=4$. However, in an individual election it is not always the case that a higher $TL$ is better for the Condorcet winner. There are seven elections where the Condorcet winner isn't the RCV winner for $TL=n-1$ but is the RCV winner for a smaller truncation level. Also, there are 21 elections with the property that there exist $i<j<k$ where the Condorcet winner is the RCV winner for $TL$ values of $i$ and $k$ but is not the RCV winner for $TL=j$. Thus, there is some Condorcet winner ``non-monotonicity'' in the data.

We note that the single-winner elections tend to produce less interesting Condorcet results than the multiwinner elections. Of the 132 single-winner elections only four (respectively one) elections satisfy that the Condorcet winner is not the RCV winner for $TL=2$ (respectively $TL=3$), and if $TL>3$ then the Condorcet winner and RCV winner always agree (although our data pool shrinks considerably for larger $TL$ values). Furthermore, we do not observe any of the non-monotonicity mentioned above in any of the single-winner elections. There are two potential reasons that we see more tame behavior in these elections. First, we simply have far fewer single-winner elections our database; second, the Condorcet dynamics of multiwinner elections might be different and more complicated.

As with the previous questions, our Condorcet results produce much less disagreement than the results of \cite{KGF}. For example, under their random model the probability that the Condorcet winner (assuming one exists) is the RCV winner with $TL=2$ is less than 80\% for each $n \in \{4,5,6\}$.

\begin{table}[]
  \centering

\begin{tabular}{l|c|c|c|c|c|c|c}
$TL$&1&2&3&4&5&6&7\\
\hline
C-useable elections &1106 & 1074 & 1011&870&669&397&207\\
\% CW wins&77.2\% & 90.2\% & 96.9\% & 98.4\% & 98.8\% & 98.5\% & 98.6\%\\
\end{tabular}
\caption{The percentage of Condorcet-useable elections in which the Condorcet winner is the RCV winner under the given $TL$.}
  \label{Cond_results}
  \end{table}

  \section{Ballot Sampling}

The results of the previous section suggest that the choice of truncation level generally has little effect on the RCV winner when using vote data from real-world elections, especially when $TL>1$. In particular, except for the uninteresting case of $n=3$, we found no elections which produce the theoretical maximum of $n-1$ different winners as $TL$ varies from 1 to $n-1$. Even though our database is large, to investigate the possible effects of ballot truncation in data like ours we use our elections to run simulations where we generate new elections by sampling ballots with replacement. This technique is common in empirical voting analyses; see \cite{RKKH} and \cite{TUK}, for example. We use this technique to investigate the worst-case scenario for the effects of ballot truncation when using elections from our database.

Before presenting our simulation results we describe the methodology. For each election in the database, we repeated the following steps 1000 times in order to generate 1000 new elections (we refer to such a new election as a \emph{pseudoprofile}).

\begin{enumerate}
\item Randomly sample min$\{$1001, number of voters$\}$ ballots from the election with replacement. 
\item Calculate the number of different RCV winners in the generated pseudoprofile as we vary $TL$ from 1 to $n-1$.
\end{enumerate}

We choose 1001 ballots as the size of our sample (except in the few elections where the number of voters is less than 1001) due to limits of computation time, and because we are interested in the worst-case scenario. While 1001 is not a small sample size, its relative smallness compared to the electorate size in some of our larger elections might result in ballot data which produces interesting outcomes.

Once the 1000 runs are complete, we record the maximum number of different winners from the 1000 generated pseudoprofiles and assign this number to the original election. We then use these numbers to build a ``worst-case'' version of Figure \ref{first_figure}, based on the maximum number of different winners observed for each election across the 1000 generated pseudoprofiles.

To see how this process plays out for a given election, we analyze a city council election in Berkeley, CA.

\begin{example}\label{Berkeley_ex}
The 2014 ranked-choice election for the City Council seat in District 8 of Berkeley, CA, contained the four candidates George Beier, Mike Cohen, Jacquelyn McCormick, and Lori Droste. In the actual election, the first round vote totals for the four candidates (in alphabetical order) were 1198, 1165, 837, and 1318. McCormick was eliminated, resulting in adjusted vote totals of 1473, 1300, and 1614 for Beier, Cohen, and Droste, repectively. After Cohen was eliminated, Droste narrowly won with 2072 votes to Cohen's 2056.

In the actual election, Droste is the RCV winner for all truncation levels. However, as the narrow vote margins suggest, if we use ballot sampling to generate a pseudoprofile then it would not be surprising to obtain an election with different winners for different truncation levels. As described above, we generated 1000 pseudoprofiles each with 1001 ballots, where the ballots are chosen at random with replacement. Of these 1000 generated elections, 614 produced only one winner, 383 produced two, and three pseudoprofiles produced three different winners. Thus, it is possible (although not likely) that vote data similar to this election could produce three different winners for the three different truncation levels. Since three is the maximum number of different winners observed, we assign a value of three to this election when constructing a new figure.

We note this election shows that using a sample of size 1001 helps create a larger maximum number of observed different winners. If we use the original electorate size then all generated pseudoprofiles produce only two different winners (Beier or Droste) for any truncation level.
\end{example}

\begin{figure}[!htb] 
\begin{center}
\includegraphics[scale=0.45]{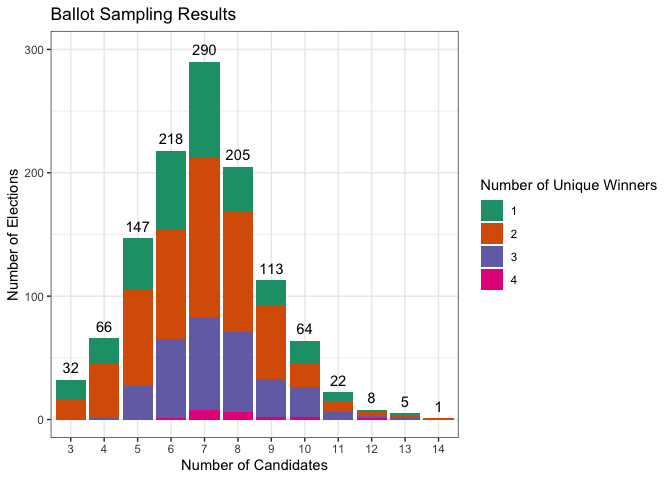}\\

\end{center}
\caption{For each election, the figure shows the maximum number of different winners as $TL$ varies from 1 to $n-1$ across the 1000 generated pseudoprofiles.}
\label{sampling_figure}
\end{figure}

The results of our ballot sampling are summarized in Figure \ref{sampling_figure}.  The complete results are available at \cite{G}. 308 elections produce only one winner across all truncation levels in every generated pseudoprofile, 546 produce a maximum of two different winners for some pseudoprofile, 297 produce a maximum of three different winners for some pseudoprofile, and 20 produce four different winners. Even with using a maximum of different winners across 1000 generated pseudoprofiles, the expected maximum number of winners across all truncation levels is only 2.02, suggesting that seeing an average of two different winners is the worst-case scenario in real-world elections. Even when using our ballot sampling methodology, we only see the theoretical maximum of $n-1$ different winners generated from a single election (Example \ref{Berkeley_ex}) with $n>3$. In general, real-world data is not likely to produce a number of different winners anywhere close to that theoretical maximum.

We again emphasize that Figure \ref{sampling_figure} represents a worst-case outcome in a number of ways. First, we use a relatively small sample of size 1001 when generating a new pseudoprofile. Second, we record only the maximum number of different winners from the 1000 runs and that maximum generally occurs infrequently, as was the case with the Berkeley election in Example \ref{Berkeley_ex}. We generated a total of 1,073,000 pseudoprofiles with five or more candidates, and only 62 pseudoprofiles (coming from 20 different elections) produced four different winners.

\begin{figure}[!hbt] 
\begin{center}
\begin{tabular}{c}
\includegraphics[scale=0.4]{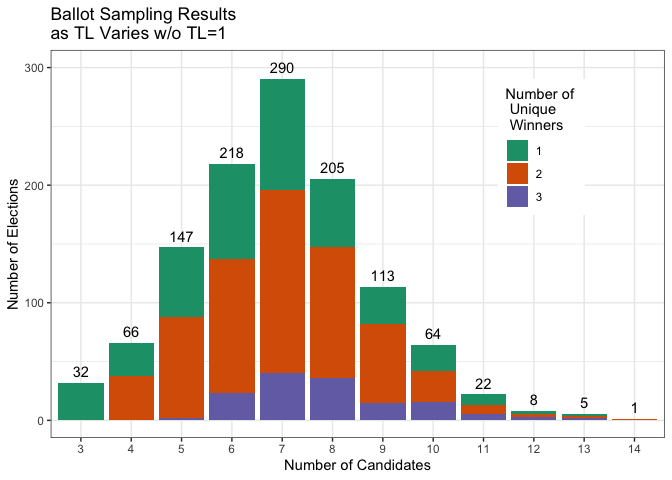}\\
\\
 \includegraphics[scale=0.4]{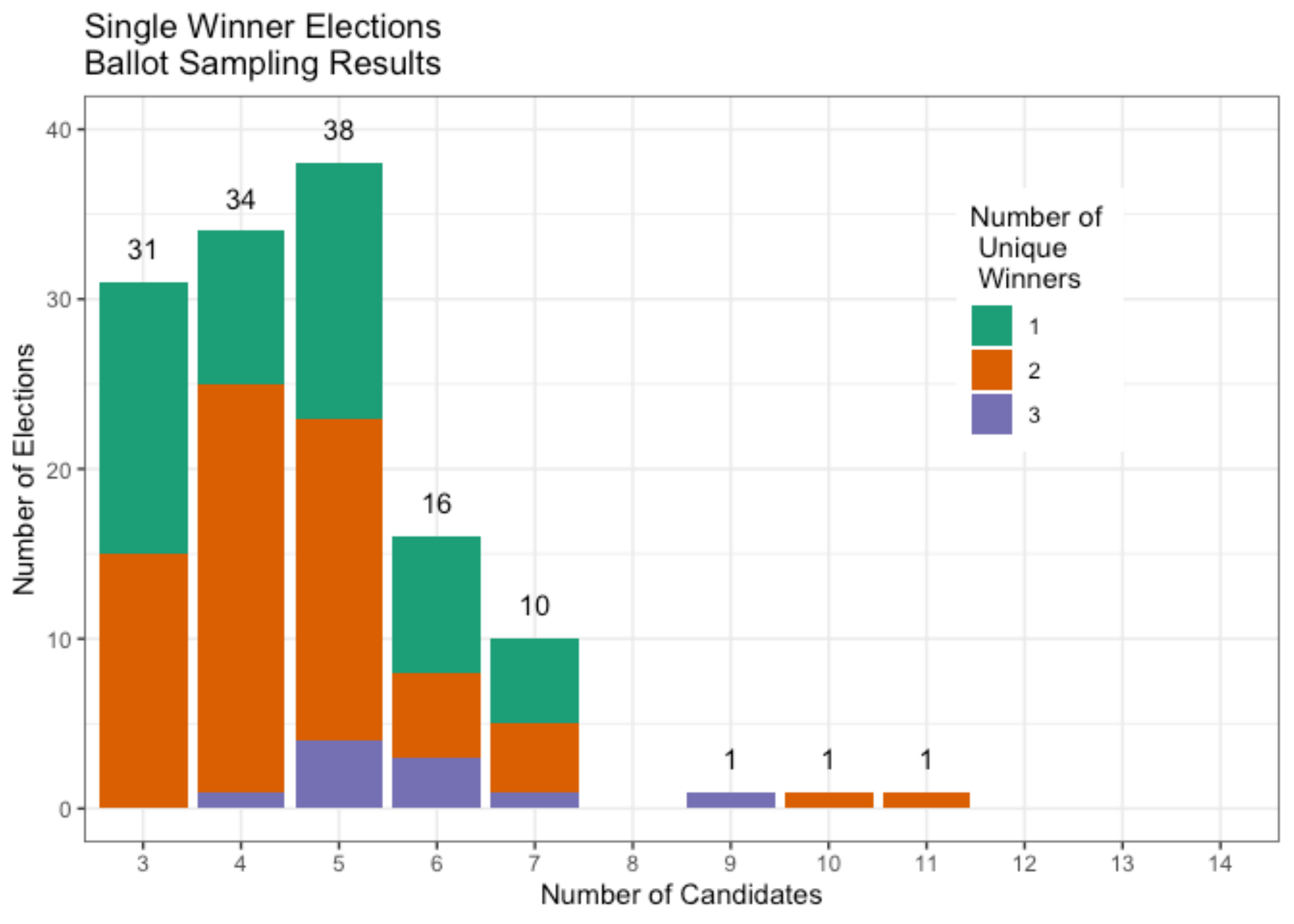} \\

\end{tabular}

\end{center}
\caption{(Top) Our sampling results, disregarding $TL=1$. (Bottom) Our sampling results for only the single-winner elections in the database.}
\label{sampling_figureb}
\end{figure}

Figure \ref{sampling_figureb} shows the sampling results when we disregard $TL=1$ (top image) and when we consider only the 132 single-winner elections in the database (bottom image). As can be seen from the figure, we only obtain four different winners in our sampling data because the plurality winner differs from the winner at all other truncation levels in the pseudoprofiles that returned four winners.  For the single-winner case, only 10 elections generated a pseudoprofile with three different winners, and this outcome was very improbable. Of the 10,000 generated pseduoprofiles from these 10 elections, 141 produced three different winners. The single-winner election which generated the most pseudprofiles with three different winners is a 2009 city council election in Aspen, CO, which generated 59 pseudoprofiles with three different winners.

\section{Conclusion}

To analyze the potential effects of truncating ballots on the winner of a ranked-choice election, we empirically investigated ballot truncation in a large database of elections. The elections we used are ideal for this analysis because they do not contain a majority candidate and voters were allowed to provide a complete preference ranking of the candidates. Assuming voters do not change their stated preferences for the top candidates on their ballot if a truncation level were imposed, our data suggests that the imposition of a truncation level greater than one rarely changes the RCV winner. Therefore, when jurisdictions such as Minneapolis or New York City impose truncation levels of 3 and 5, respectively, the effect on the RCV winner likely manifests only insofar as  voters choose to provide an insincere ranking of their top candidates in response to the truncation level. The extent to which voters behave in this way, and the question of whether such voter adaptation is undesirable, are beyond the scope of this article and represent fertile ground for future research.

{\footnotesize

\end{document}